
\magnification\magstep1
\hsize 15.5truecm
\scrollmode


\def\st{structure }
\def\sts{structures }

\def\nd{\noindent}

\overfullrule=0pt

\def\cf.{{\sl cf.}}

\def\pr{\partial }
\def\mo{morphisme}

\def\ha{Hamiltonian}
\def\pos{Poisson structure }
\def\poss{Poisson structures }

\def\man{manifold}
\hbox{}

 at 17.28truept
 at 14.4truept
 at 12truept
 at 10.95truept
 at 17.28truept
 at 10truept

\def\title#1{%
\vskip0pt plus.3\vsize\penalty-100%
\vskip0pt plus-.3\vsize\bigskip\vskip\parskip%
\bigbreak\bigbreak\centerline{\bf #1}\bigskip%
}

\def\chapter#1#2{\vfill\eject
\centerline{\bf Chapter #1}
\vskip 6truept%
\centerline{\bf #2}%
\vskip 2 true cm}

\def\section#1#2{%
\def\\{#2}%
\vskip0pt plus.3\vsize\penalty-100%
\vskip0pt plus-.3\vsize\bigskip\vskip\parskip%
\par\noindent{\bf #1\hskip 6truept%
\ifx\empty\\{\relax}\else{\bf #2\smallskip}\fi}}

\def\subsection#1#2{%
\def\\{#2}%
\vskip0pt plus.3\vsize\penalty-20%
\vskip0pt plus-.3\vsize\medskip\vskip\parskip%
\def\TEST{#1}%
\noindent{\ifx\TEST\empty\relax\else\bf #1\hskip 6truept\fi%
\ifx\empty\\{\relax}\else{#2\smallskip}\fi}}

\def\proclaim#1{\medbreak\begingroup\noindent{\bf #1.---}\enspace\sl}

\def\endproclaim{\endgroup\par\medbreak}

\def\qbox#1{\quad\hbox{#1}}

\def\lbreak{\hfil\break}


\def\comfig#1#2\par{
\medskip
\centerline{\hbox{\hsize=10cm\eightpoint\baselineskip=10pt
\vbox{\noindent #1}}}\par\centerline{ Figure #2}}

\def\figcom#1#2\par{
\medskip
\centerline
{Figure #1}
\par\centerline{\hbox{\hsize=10cm\eightpoint\baselineskip=10pt
\vbox{\noindent #2}}}}


\def\lbreak{\hfill\break}

\def\comfig#1#2\par{
\medskip
\centerline{\hbox{\hsize=10cm\eightpoint\baselineskip=10pt
\vbox{\noindent{\sl  #1}}}}\par\centerline{{\bf Figure #2}}}

\def\figcom#1#2\par{
\medskip
\centerline
{{\bf Figure #1}}
\par\centerline{\hbox{\hsize=10cm\eightpoint\baselineskip=10pt
\vbox{\noindent{\sl  #2}}}}}

\def\em{\sl}

\def\\{\hfill\break}

\def\bibitem{\item}

\def\lbreak{\hfill\break}


\def\pr {\partial }

\def\nd{\noindent}

\def\st{structure }

\def\a{\alpha}

\def\d{\delta}

\def\e{\varepsilon}

\def\la{\lambda}
\def\La{\Lambda}

\def\f{\varphi}

\def\CC{{\bf C}}

\def\RR{{\bf R}}

\def\cL{{\cal L}}


\def\la{\lambda}

\def\CI{C^\infty}

\def\sqr#1#2{{\vcenter{\hrule height.#2pt%
\hbox{\vrule width.#2pt height#1pt\kern#1pt%
\vrule width.#2pt}%
\hrule height.#2pt}}}

\title{ Higher Gel'fand-Dikii structures}

\centerline{{\rm B. Enriquez, A. Orlov and V. Rubtsov}}

\medskip
\nd {\bf Abstract.} {\em
We apply the procedure of Magri and Weinstein to produce an infinity
of compatible Poisson structures on a bihamiltonian manifold, to the case of
the KdV phase space. The higher Gel'fand-Dikii structures thus obtained
contain non local terms, which we express with the help of the r.h.s. of the
KdV hierarchy. We also give a generating function for all these Poisson
structues, in terms of the Baker-Akhiezer functions. Finally we describe the
symplectic leaves of these Poisson structures.}

\section{1.}{Definition of the higher Gel'fand-Dikii structures.}

In [M], [W], it is explained how to associate to a pair of \poss \ $V_{i}$,
$i=1,2$, on a \man \ $P$, one of which $(V_{1})$ is symplectic, an infinite
sequence $V_{n}$ of compatible Poisson structures. The maps
$V_{i}: T^{*}P\to TP$ allow to
define the Nijenhuis operator $\La$~: $TP\to TP$ by $V_{2}V_{1}^{-1}$~; we
then
pose $V_{n}=\La^{n-1}V_{1}$.

A natural example of compatible \poss is provided by the first Gel'fand-Dikii
and second (Adler-)Gel'fand-Dikii (GD1 and GD2) structures on \man s of
differential
operators. Remark that GD1 is not exactly symplectic, since if we define
the \man \ to be $\cL_{2}=\{\pr^{2}+u, \, u\in\CI_{c}(\RR)\}$, then
$T_{L}\cL_{2}=\{\d u, \, \d u\in\CI_{c}(\RR)\}$ (the subscript $c$ means
compactly supported functions) for $L\in\cL_{2}$, and $T^{*}_{L}\cL_{2}$ is
the set of forms given by $\d u\mapsto\int_{\RR}\xi \d u,\,
\xi\in\CI_{c}(\RR)$. The map $V_{1}$ is then $T^{*}_{L}\cL_{2}\to
T_{L}\cL_{2}\, ,\, \xi\mapsto\xi'$~; it has a cokernel of dimension one.

A natural way to overcome this difficulty is to enlarge the space of
functionals on $\cL_{2}$ (recall that in the usual formalism it is the set
of maps $u\mapsto\int_{\RR}\, f(x,u(x),\cdots,u^{(k)}(x))dx$, $f$ smooth with
compact support in $x$ and polynomial in the other variables).

We now pose Fun $\cL_{2}$ to be the linear span of the functionals
$$
u\mapsto \int_{\RR^{k}}f_{1}(x_{1},u^{(\a)}(x_{1}))\cdots f_{k}(x_{k},
u^{(\a)}(x_{k}))\prod^{N}_{\a=1}\,
\e(x_{i_{\a}}-x_{j_{\a}})\prod^{k}_{i=1}dx_{i}\, ,\, i_{\a}< j_{\a}\, ,
$$
with $\cup^{k-1}_{\a=1}\, \{i_{\a},j_{\a}\}=\{1,\cdots,k\}$, and $f_{k}$
polynomial in the $u^{(\a)}(x_{k})$, $\a\ge 0$ with coefficients $\CI_{c}$
functions of $x_{k}$~; $\e$ is the Heaviside function ($\e(x)=-{1\over 2}$
if $x<0\, ,\, {1\over 2}$ else)~; $N$ is arbitrary.

We still pose $T_{L}\cL_{2}=\{\d u,\, \d u\in\CI_{c}(\RR)\}$ but define now
$T^{*}_{L}\cL_{2}$ as the set of forms given by $\d u\mapsto\int_{\RR}
\tilde\xi\d
u$, $\tilde\xi$ a smooth function on $\RR$ with opposite limits at
$+\infty$ and
$-\infty$. The map $V_{1}: T^{*}_{L}\cL_{2}\to T_{L}\cL_{2}$,
$\tilde\xi\mapsto\tilde\xi'$ is now a linear iso\mo \ and we can try to
apply the
Magri-Weinstein procedure.

Let us consider in $T^{*}_{L}\cL_2$ the subspace $T^{*}_{0}$ linearly
spanned by the
functions
$$
x_{1}\mapsto\int_{\RR^{k-1}}f_{1}(x_{1},u^{(\a)}(x_{1}))\dots
f_{k}(x_{k},u^{(\a)}(x_{k}))\prod^{N}_{\a=1}\e(x_{i_{\a}}-x_{j_{\a}})
\prod^{k}_{i=2}dx_{i}\, ,
$$
where exactly one $i_{\a}$ or $j_{\alpha}$ is equal to one and
$f_{1},\dots,f_{k}$ are as
above, but the constant term of $f_{1}$ is of the form constant $+\CI_{c}$
($N$
is arbitrary); and in $T_{L}\cL_{2}$ the subspace $T_{0}$ linearly spanned by
analogous expressions, with the conditions that the constant terms of the
polynomial $f_{i}$
are $\CI_{c}$ (and the other are smooth), and no restriction on $(i_{\a})$,
$(j_{\a})$.
\lbreak
The map $V_{1}$ induces a linear isomorphism between $T^{*}_{0}$ and
$T_{0}$~:
indeed, $V_{1}(T_{0}^{*})\subset T_{0}$, and the preimage of the element of
$T_{0}$
$$x_{1}\mapsto\int_{\RR^{k-1}}f_{1}(x_{1},u^{(\a)}(x_{1}))\dots
f_{k}(x_{k},u^{(\a)}(x_{k}))\prod^{N}_{\a=1}\e(x_{i_{\a}}-x_{j_{\a}})
\prod^{k}_{i=2}dx_{i}\, ,
$$
is
$$
x_{0}\mapsto\int_{\RR^{k}}1(x_{0})f_{1}(x_{1},u^{\a}(x_{1}))\dots
f_{k}(x_{k},u^{(\a)}(x_{k}))\cdot\e(x-x_{1})-\prod^{N}_{\a=1}\e(x_{i_{\a}}
-x_{j_{\a}})\prod^{k}_{i=1}dx_{i}
$$
which belongs to $T^{*}_{0}$ ($1(x_{0})$ is the constant function of $x_{0}$
equal to 1). On the other hand, $V_{2}(T^{*}_{0})\subset T_{0}$ (recall that
$V_{2}={1\over 4}(\pr^{3}+4u\pr+2u_{x})$). Note also that $V_1$ and $V_2$
define antisymmetric bilinear forms on $T^*_0$ (or more generally on the
space
$T^*(-1)$, if $T^*(c)=\{\xi\in C^{\infty}(\RR)|\xi$ is constant at both
infinities and $\rm{lim}_{+\infty}\xi=
\sl{c}\rm{lim}_{-\infty}\xi\}$; if $\xi,\eta\in T^*(c)$,
$\langle V_1(\xi),\eta\rangle+\langle \xi,V_1(\eta)\rangle=(1-c^2)
\rm{lim}_{+\infty}\xi\rm{lim}_{+\infty}\eta$; and $V_1$ is not bijective
from
$T^*(1)$ to $\CI_c(\RR)$. On the other hand, $V_2$ defines an antisymmetric
form
 on any $T^*(c)$.) We can check that $V_1$ and $V_2$
still define Poisson structures
on Fun $\cL_2$.
\lbreak
The recursion operator $\La$ is then well defined from $T_{0}$ to itself~; we
get mappings $V_{n}=\La^{n}\pr$ from $T_{0}^{*}$ to $T_{0}$.

If now $F$ belongs to Fun $\cL_{2}$, $dF$ belongs to $T^{*}_{0}$, and
$V_{n}(dF)$ to $T_{0}$~; let then $G$ be an other function of Fun $\cL_{2}$.
 It
is easy to see that $V_{n}(dF)G$ still belongs to Fun $\cL_{2}$. Thus, the
$V_{n}$ define bilinear operations on Fun $\cL_{n}$. The Magri-Weinstein
arguments
allow to show that they form an infinite family of compatible Poisson
brackets,
that we will call higher Gel'fand-Dikii brackets. We will now analyse these
brackets more closely.

\section{2.}{Non local part of the higher Gel'fand-Dikii structures.}
 Recall that on the KdV phase space $\cL_{2}=\{\pr^{2}+u,\,
u\in\CI_{c}(\RR)\}$, GD1
is given by the operator $V_{1}=\pr$ and GD2 by $V_{2}={1\over
4}(\pr^{3}+2u\pr+2u_{x})$, leading to the brackets
$$
\{u(x),u(y)\}_{1}=\d'(x-y), \, \{u(x), u(y)\}_{2}={1\over
4}\d'''(x-y)+{1\over 2}(u(x)+u(y))\d'(x-y)\, .
$$
The recursion operator is $\La={1\over 4}(\pr^{2}+4u+2u_{x}\pr^{-1})$. The
operator of the $n$-th GD \st is then $\La^{n}\pr$. The result of this
section
is that
$\La^{n}\pr$ can be written (local part) +
$\mathop{\sum}\limits^{n}_{i=1}K'_{i}\pr^{-1}K_{n-i}\pr$, where local part
is a
differential operator (with coefficients differential polynomials in $u$),
and
where $K_{n}$ are the right hand sides of the KdV hierarchy~:
$K_{n+1}=\La^{*}K_{n}$ (we set $(f\pr^{k})^{*}=(-\pr)^{k}f$, $f$ a function,
$k$ an integer), $K_{0}=1$; $K_n$ is then a polynomial differential in $u$,
containing (by assumption) no constant term if $n\ne 0$.

\nd
We first prove some relations in the algebra of formal pseudodifferential
operators~:
\proclaim{Lemma} If $a,b, a_{1},b_{1}$ are differential polynomials in $u$
and
$P$ is a differential operator, we have
$$
\eqalign{
P(a\pr^{-1}b)&= [Pa]\pr^{-1}b+R\, ,\cr
(a\pr^{-1}b)P&=a\pr^{-1}[P^{*}b]+S\, \cr
(a\pr^{-1}b)(a_{1}\pr^{-1}b_{1})&=[a\pr^{-1}ba_{1}]\pr^{-1}b_{1}+a\pr^{-1}
[(a_{1}\pr^{-1}b_{1})^{*}b]\cr}
$$

\nd
where e.g. $[Pa]$ denotes the function obtained from the operation of $P$ on
$a$, and $R$ and $S$ are differential operators (in the last relation,
$ba_{1}$ as assumed to be a total derivative $c'$ and $[\pr^{-1}ba_{1}]=c$ in
the r.h.s.).
\endproclaim

The first relation is proved by examining the cases $P$ = function and
$P=\pr$,
and by the remark that the set of $P$ satisfying it forms an algebra~; the
second relation can be deduced from it by applying $*$. To obtain the third
one, we replace
in the r.h.s. $ba_{1}$ by $\pr[\pr^{-1}ba_{1}]-[\pr^{-1}ba_{1}]\pr$.

We apply this lemma to the computation of the non local part of $\La^{n}$.
$\La = $
(local part) + $K'_{1}\pr^{-1}$ since $K_{1}={u\over 2}$. Assume that
$\La^{n}$
is written $P_{n}+\mathop{\sum}\limits^{n}_{i=1}K'_{i}\pr^{-1}K_{n-i}$,
$P_{n}$
a differential operator. Then
$$
\eqalign{
\La^{n+1} &= \hbox{differential operator}+{1\over 4}
[P_{n}.2u_{x}]\pr^{-1}\cr
& \quad  +\mathop{\sum}\limits^{n}_{i=1}K'_{i}\pr^{-1}\left[{1\over
4}(\pr^{2}+4u)^{*}K_{n-i}\right]+
\mathop{\sum}\limits^{n}_{i=1}K'_{i}\left[{\pr^{-1}\over
4}K_{n-i}2u_{x}\right]\pr^{-1}\cr
& \qquad -{1\over
4}\mathop{\sum}\limits^{n}_{i=1}K'_{i}\pr^{-1}[\pr^{-1}2u_{x}K_{n-i}]\cr
&= [\La^{n}{u_{x}\over
2}]\pr^{-1}+\mathop{\sum}\limits^{n}_{i=1}K'_{i}\pr^{-1}K_{n+1-i}+ \hbox
{diff. op.}\cr
&= K'_{n+1}\pr^{-1}+\mathop{\sum}\limits^{n}_{i=1}K'_{i}\pr^{-1}K_{n+1-i}+
\hbox{diff. op.}\cr}
$$
since from $\La^{*}=\pr^{-1}\La \pr$ follows $K'_{n+1}=\La K'_{n}$ and so
$K'_{n+1}=\La^{n}K'_{1}=\La^{n}{u_{x}\over 2}$. From $K_{s+1}=\Lambda^*K_s$
follows that $2u_xK_s$ is the total derivative of a differential polynomial
in
$u$; in $[\pr^{-1}2u_xK_s]$ we fix the constant term of this polynomial to
zero.
 \lbreak
This establishes $\La^{n} = \hbox{local
part}+\mathop{\sum}\limits^{n}_{i=1}K'_{i}\pr^{-1}K_{n-i}$ by induction.
Our result follows.

\nd
The  Poisson bracket for GDn is then~:
$$
\eqalign{
\{u(x),u(y)\}_{n} &=\hbox{(local part)}+
\mathop{\sum}\limits^{n-1}_{i=1}K'_{i}(x)\pr^{-1}_{x}K_{n-1-i}(x)\pr_{x}
\d(x-y)\cr
&=\hbox{(local part)}-
\mathop{\sum}\limits^{n-2}_{i=1}K'_{i}(x)K'_{n-1-i}(y)\e(x-y)\, ,\cr}
$$

\nd
$\hbox{(local part)} = {1\over 4^{n-1}}\d^{(2n-1)}(x-y)+
\mathop{\sum}\limits^{2n-2}_{i=0}p_{i}(u)(x)\d^{i}(x-y)$, $p_{i}$ being
differential polynomials in $u$.\lbreak

\section{3. A generating function for the higher GD structures.}{}
Let $\psi_{\la}$ and $\psi^*_{\la}$ be conjugated wave (Baker-Akhiezer)
functions for $\pr^2+u$. That is, $(\pr^2+u)\psi_{\la}=\la^2\psi_{\la}$,
$(\pr^2+u)\psi^*_{\la}=\la^2\psi^*_{\la}$, $\psi_{\la}(x)=e^{\la x}(1+
\sum_{i\ge 1}u_i(x)\la^{-i})$, $\psi^*_{\la}(x)=e^{-\la x}(1+
\sum_{i\ge 1}u_i(x)(-\la)^{-i})$, $u_i(x)=0$ for $x$ large negative enough.
Pose $R_+=\psi_{\la}^2$, $R_0=\psi_{\la}\psi_{\la}^*$, $R_-=\psi_{\la}^{*2}$.
The $R_i$ satisfy ${1\over 4}(\pr^3+4u\pr+2u_x)R_i=\pr R_i$.
Then we have the identity
$${\la^2\over{\La-\la^2}}={1\over 2}R'_+\pr^{-1}R_-
+{1\over 2}R'_-\pr^{-1}R_+
-R'_0\pr^{-1}R_0
 \ \ \ \ \ (*)
$$
It follows from
$$\eqalign{
{1\over 4}(\pr^2+4(u-\la^2)+2u_x\pr^{-1}){1\over 2}R'_+\pr^{-1}R_-&=
          {1\over 2}R'_+({1\over 4}\pr^2+u-\la^2)\pr^{-1}R_-+
             {1\over 8}(2 R_+''+R_+''')\pr^{-1}R_-\cr
        &\ \ \ \ \ \  \ \ \ \ \ \
+{u_x\over 4}R_+\pr^{-1}R_--{u_x\over 4}\pr^{-1}R_+R_-\cr
    &= {1\over 8}\pr R'_+R_-+{1\over 4}R_+''R_- -{u_x\over 4}\pr^{-1}R_+R_-;
\cr}
$$
after summation of the two other terms, the last expression gives zero.
We then have
$$\eqalign{(\La-\la^2)({1\over 2}R'_+\pr^{-1}R_- +{1\over 2}R'_-\pr^{-1}R_+
-R'_0\pr^{-1}R_0)&=
   {1\over 8}(R''_+R_-+R''_-R_+-2R''_0R_0)\cr
            &= {1\over 4}(R_0^{\prime 2}-R'_+R'_-)\cr
            &= {1\over 4}(\psi_{\la}\psi^{*\prime}_{\la}-\psi'_{\la}
\psi^{*}_{\la})^2=\la^2\cr}$$
(the Wronskian of $\psi_\la$ and $\psi^*_{\la}$ is constant, and takes the
value $2\la$ at $-\infty$).

{}From $(*)$ follows a formula generating all the $\La^k\pr$:
$$
-\sum_{k\ge 0}{{\La^k\pr}\over{\la^{2k}}}=-{1\over 2} R'_+\pr^{-1} R'_-
-{1\over 2} R'_-\pr^{-1} R'_+
+R'_0\pr^{-1}R'_0
$$
(we have used $R_+R_-=R_0^2$). Posing $R_{\pm}=e^{\pm2\la x}\hat{R}_{\pm}$,
we have $R'_{\pm}=e^{\pm2\la x}(\hat{R}'_{\pm}\pm2\la\hat{R}_{\pm}$,
and so
$$\eqalign{
-\sum_{k\ge 0}{{\La^k\pr}\over{\la^{2k}}}=-{1\over 2}
(\hat{R}'_++2\la\hat{R}_+)&
(\pr-2\la)^{-1}(\hat{R}'_- -2\la\hat{R}_-) \cr &
-{1\over 2} (\hat{R}'_- -2\la\hat{R}_-)(\pr+2\la)^{-1}
(\hat{R}'_++2\la\hat{R}_+)
+R'_0\pr^{-1}R'_0
}$$
Here $(\pr\pm2\la)^{-1}$ should be expanded as $\sum_{k\ge 0}{\pr^{k}\over
{(\pm 2\la)^{k+1}}}$. So the only non local contribution is the one of
the terms in $R_0$, which enables to recover the results of the last section
(since $R_0=1+\sum_{n\ge 1}{K_n\over{\la^{2n}}}$). Remark also that all the
expressions involved in the expansion in $\la$ are polynomial differentials
in
$u$ since they are invariant under the transformations $\psi_{\la}(x)\mapsto
c(\la)\psi_{\la}(x)$,
$\psi^*_{\la}(x)\mapsto c(\la)^{-1}\psi^*_{\la}(x)$, $c(\la)\in
\CC[[\la^{-1}]]^*$.

If $w$ is then a primitive of the variable $u$, we get
$$\eqalign{
\{w(x),w(y)\}_{\la}&=\sum_{n\ge 0}{\{w(x),w(y)\}_n\over{\la^{2n}}}\cr
    &=({1\over 2}R_+(x)R_-(y)+{1\over 2}R_-(x)R_+(y)-R_0(x)R_0(y))
\varepsilon(x-y)
}$$
Here $e^{\pm2\la(x-y)}\varepsilon(x-y)$ should be expanded as $-\sum_{k\ge 0}
{{\delta^{(k)}(x-y)}\over{(\pm2\la)^{k+1}}}$.

In the case of GD3, the bracket is~:
$$
\eqalign{
\{u(x),u(y)\}_{3}&= {1\over 16}\, \d^{V}(x-y)+{1\over 4}\,
(u(x)+u(y))\d'''(x-y)\cr
&\quad +{1\over 8}(u'(x)-u'(y))\d''(x-y)+{1\over
2}(u(x)^{2}+u(y)^{2})\d'(x-y)\cr
&\qquad -{1\over 4}\, u'(x)u'(y)\e(x-y)\, .\cr}
$$

For this third GD \st the functional $\int_{\RR}u(x)dx$ is a \ha \ for
the KdV
equation. It is not clear what the \ha s for the KdV equations are in the
higher GD \sts (it should be non local quantities related to the solutions of
$(\pr^{2}+u+\la)\f=0$ for $\la$ small).

\section{Remarks.}{}

1. $R_+$ and $R_-$ can also be used to give a generating function for
``$A$-operators'' for $\La$: $${{\pr\La}\over{\pr t}}=\sum_{n\ge 0}
{1\over{\la^n}}
{{\pr\La}\over{\pr t_n}}=[\La,R_+'\pr^{-1}R_--R_-'\pr^{-1}R_+],
$$
$t_i$ being the times of the KdV hierarchy.

2. In the last section,  we used the fact that $u_xK_n$ is a total
derivative.
We can show more generally that for any $i$ and $j$, $K_i'K_j$ is a total
derivative. Recalling that $R_0=1+\sum_{n\ge 1}{K_n\over{\la^{2n}}}$, it is
enough to prove that $R_0(\la,x)R'_0(\mu,x)$ is a total derivative. Indeed
(noting $W(f,g)=fg'-f'g$),
$$
(W(\psi_{\la},\psi_{\mu}^*)
W(\psi_{\la}^*,\psi_{\mu}))'=(\mu^2-\la^2)(R_0(\la)R'_0(\mu)-R_0'(\la)
R_0(\mu)),
$$
so that a primitive of $R_0(\la)R'_0(\mu)$ is
$${1\over 2}(
{1\over{\mu^2-\la^2}}W(\psi_{\la},\psi_{\mu}^*)W(\psi_{\la}^*,\psi_{\mu})+
R_0(\la)R_0(\mu));
$$
 it consists of differential polynomials in $u$ since it
is invariant by the transformations $\psi_\la(x)\mapsto c(\la)\psi_\la(x)$,
$\psi^*_\la(x)\mapsto c(\la)^{-1}\psi^*_\la(x)$,
$\psi_\mu(x)\mapsto d(\la)\psi_\mu(x)$,
$\psi^*_\mu(x)\mapsto d(\la)^{-1}\psi^*_\mu(x)$,
$c(\la)\in\CC[[\la^{-1}]]^*$, $d(\mu)\in\CC[[\mu^{-1}]]^*$.

\section{3.}{Symplectic leaves of the higher GD structures.}

Let $M$ be a finite dimensional manifold,  endowed with a
symplectic \st $V_{0}: T^{*}M\to TM$ and a \pos $V_{1}: T^{*}M\to TM$,
which we
assume to be compatible. Let $\La=V_{1}V_{0}^{-1}$ be the recursion
operator.\lbreak
Let $P(\La)=\mathop{\prod}\limits^{n}_{i=1}(\La-\la_{i})^{k_{i}}$ be an
arbitrary
monic polynomial (the $\la_{i}$ are all different, $n\ge 1$). Then the
general
form of the
\poss defined in [W] is $V_{P}=P(\La)V_{0}$.

Let us analyse the symplectic leaves of $V_{P}$. They are the integral
manifolds of the forms $\xi$ such that $P(\La) V_{0}\xi
=\mathop{\prod}\limits^{n}_{i=1}
(\La-\la_{i})^{k_{i}}V_{0}\xi=0$. Any $\xi$ satisfying this relation is such
that at any point $x$ of $M$, $\xi(x)$ belongs to the sum $\sum_{i=1}^n
D_i(x)$
of the subspaces $D_i(x)$ of $T^*_xM$ consisting of the forms satisfying
$(\Lambda-\lambda_i)^{k_i}V_0\xi=0$. So the tangent vectors to the
symplectic
leaves of $V_P$ are exactly the vectors tangent to the symplectic leaves of
all
the $V_{(\Lambda-\lambda_i)^{k_i}}$, and the symplectic leaves of $V_{P}$ are
the intersections of the symplectic leaves of the
$V_{(\La-\la_{i})^{k_{i}}}$.

Let us apply this discussion to the \sts on the KdV phase space discussed
above. First we describe the symplectic leaves of $V_{\La^{k}V_{0}}\, (k\geq
1)$.\lbreak
For this, we solve $\La^{k}V_{0}\xi=0$, $\xi$ a covector at $\pr^{2}+u$.
Recall
([K])
that the Casimir functions of $V_{1}$ are the functions of tr $M(u)$ ($M(u)$
 is
the monodromy operator of $\pr^{2}+u$), so that $\La V_{0}\xi=0$ means that
$\xi$
is proportional to $d$tr $M(u)$. Let us solve $\La^{2}V_{0}\xi=0$.

\nd
Differentiating  $(\La+\la)V_{0}d \, {\rm tr}\, M(u+\la)=0$, we get
$V_{0}d \, {\rm tr}\,
M(u)+\La V_{0}d\pr_{\la}\, {\rm tr}\, M(u+\la)|_{\la=0}=0$.\lbreak
$\La^{2}V_{0}\xi=0$ means that $\La V_{0}\xi$ proportional to
$d\, {\rm tr}\,
M(u)
$, i.e. $V_{0}\xi$ is a linear combination of $d\, {\rm tr}\, M(u+\la)$  and
$d\pr_{\la}\, {\rm tr}\, M(u+\la)|_{\la=0}$. In the same way, we see that
$\La^{k}V_{0}\xi=0$ has for solutions the linear combinations of
$d\, {\rm tr}\, M(u)$, $d\pr_{\la}\, {\rm tr}\,M(u+\la)|_{\la=0},\dots,
d\pr_{\la^{k-1}}\, {\rm tr}\, M(u+\la)|_{\la=0}$.

\nd
Hence the symplectic leaves of $\La^{k}V_{0}$ are the \man s
$$
{\rm tr}\, M(u)=C_{0},\dots, \pr_{\la}^{k-1}\, {\rm tr}\,
M(u+\la)|_{\la=0}=C_{k-1}\, ,$$
$C_{0},\dots,C_{k-1}$ are constants.
In Miura coordinates $[\pr^{2}+u=(\pr+\f')(\pr-\f'),\,  \f(-\infty)=0]$ these
conditions can be written

$$(e^{\f}+e^{-\f})(+\infty)=C_{0},$$

$$
\eqalign{
e^{\f(\infty)}\int_{-\infty}^{\infty}dx e^{-2\f(x)}\int^{x}_{-\infty}e^{2\f}
+e^{-\f(\infty)}\int^{\infty}_{-\infty}dx
e^{2\f(x)}\int^{x}_{-\infty}e^{-2\f}
&=C_{1}\,
,\cr
e^{\f(\infty)}\int_{-\infty}^{\infty}dx e^{-2\f(x)}\int^{x}_{\infty}dy
e^{2\f(y)}
\int^{y}_{-\infty}dz e^{-2\f(z)}\int^{z}_{-\infty}e^{2\f}\cr
+e^{-\f(\infty)}\int_{-\infty}^{\infty}dx e^{2\f(x)}\int^{x}_{-\infty}
dy e^{-2\f(y)}
\int^{y}_{-\infty}dz e^{2\f(z)}\int^{z}_{-\infty}e^{-2\f}&=C_{2}\,
 , \dots\cr}
$$

\section{Remark.}{} It could be interesting to compute the Poisson brackets
between the monodromy operators of the $\pr^2+u+\la$, with Poisson brackets
 of
$u$ corresponding to the higher GD structures. In the case of GD1 the
Poisson
brackets between monodromies are given by the rational $r$-matrix $t\over
{\la-\mu}$ on the group $SL_2(\CC[[\la]])$ (Faddeev-Takhtajan),
in the case of GD2 they are given by the trigomometric $r$-matrix
${1\over 2}{{\la+\mu}\over{\la-\mu}}t+r$ (in the Belavin-Drinfeld
notations). These two Poisson structures on $SL_2(\CC[[\la]])$ are
compatible.

\section{Acknowledgments.}{} This work was done while one of us (B.E.) was
visiting the ITEP (Moscow), and he (B.E.) would like to thank
heartily D. Lebedev for his kind hospitality.

\section{References.}{}

\bibitem{[A] }  M. Adler, {\it On a trace functional for formal
 pseudodifferential operators and
the symplectic \st of the {\rm K-dV} equation}, Invent. Math. {\bf 50}
(1979),
 219-248.

\bibitem{[GD] } I.M. Gel'fand, L.A. Dikii, {\it Family of Hamiltonian \sts
connected with
integrable nonlinear equations}, preprint IPM (Moscow) 136, in russian,
(1979).

\bibitem{[K] } A.A. Kirillov, {\it Infinite dimensional Lie groups: their
orbits, invariants and representations. Geometry of moments}, Lect. Notes in
Math., vol. 970, 101-123, Springer, 1982.

\bibitem{[M] } F. Magri, {\it A geometric approach to the nonlinear solvable
equations}, Lect. Notes in Physics, vol. 120, 233-263, Springer, 1980.

\bibitem{[W] } A. Weinstein, {\it Poisson \sts \ and Lie algebras}, Soc.
Math.
 Fr.,
Ast\'{e}risque, hors s\'{e}rie (1985), 421-434.

$$
\matrix{
\hfil \hskip 8truecm &\qbox{B. Enriquez}\hfill\cr
\hfil \hskip 8truecm &\qbox{Centre de Math\'{e}matiques}\hfill\cr
\hfil \hskip 8truecm &\qbox{URA 169 du CNRS}\hfill\cr
\hfil \hskip 8truecm &\qbox{Ecole Polytechnique}\hfill\cr
\hfil \hskip 8truecm &\qbox{91128 Palaiseau}\hfill\cr
\hfil \hskip 8truecm &\qbox{France}\hfill\cr
\hfil \hskip 8truecm &\qbox{ }\hfill\cr
\hfil \hskip 8truecm &\qbox{A. Orlov}\hfill\cr
\hfil \hskip 8truecm &\qbox{ul. Krasikova 23,}\hfill\cr
\hfil \hskip 8truecm &\qbox{Institute of Oceanology,}\hfill\cr
\hfil \hskip 8truecm &\qbox{117218 Moscow}\hfill\cr
\hfil \hskip 8truecm &\qbox{Russia}\hfill\cr
\hfil \hskip 8truecm &\qbox{ }\hfill\cr
\hfil \hskip 8truecm &\qbox{V. Rubtsov}\hfill\cr
\hfil \hskip 8truecm &\qbox{5/1, apt. 68,}\hfill\cr
\hfil \hskip 8truecm &\qbox{60th anniversary of October av.,}\hfill\cr
\hfil \hskip 8truecm &\qbox{117334 Moscow}\hfill\cr
\hfil \hskip 8truecm &\qbox{Russia}\hfill\cr}
$$

\bye